# White Paper: A Method for Comparing Hedge Funds

Uri Kartoun[1]

*2501 Porter St. #806, Washington D.C., 20008, U.S.A.*

*Stockato LLC*

+1-202-374-4007

uri@stockato.com

**Abstract.** The paper presents new machine learning methods: *signal composition*, which classifies time-series regardless of length, type, and quantity; and *self-labeling*, a supervised-learning enhancement. The paper describes further the implementation of the methods on a financial search engine system to identify behavioral similarities among time-series representing monthly returns of 11,312 hedge funds operated during approximately one decade (2000 - 2010). The presented approach of cross-category and cross-location classification assists the investor to identify alternative investments.

## 1. Introduction

STANDARD filtering menus, checkboxes, and textboxes are interface controls commonly used in hedge funds' selection database systems. Such screening methods aim to save time by narrowing one's search to a manageable number of specific investments for further research and examination. One disadvantage of these classification methods is the requirement of the user to be financially knowledgeable and provide the comparison criteria. Criteria include parameters such as performance history, investment style and category, and fees, to name a few. Another disadvantage is the complexity of the user interfaces asking the user to provide parameters through the variety of available interface controls. Another limitation is the lack of ability to classify hedge funds based on similarities in behavioral patterns. An example of a behavioral pattern would be a time-series considered in a specific time period, wherein the time-series is a sequence of data points that represent the monthly returns of the hedge fund.

---

[1] Uri Kartoun was with the Department of Industrial Engineering and Management, Ben-Gurion University of the Negev, Beer-Sheva 84105, ISRAEL, and with Microsoft Corporation, One Microsoft Way, Redmond, WA, U.S.A. He is now with Stockato LLC, 2501 Porter St. #806, Washington D.C., 20008, U.S.A. (phone: +1-202-374-4007; e-mail: uri@stockato.com).

Related work addressing classification of hedge funds includes, for example, Das, 2003, who describes experiments that involve clustering algorithms. Similar computational methods applied on hedge funds are described by Liang, 2005. *Morningstar, Inc.*, supports 31 hedge fund categories, which map into six broad category groupings (directional equity, relative value, directional debt, global/derivatives, event, and multi-strategy) as described in *The Morningstar Category Classifications for Hedge Funds*, 2012.

Previous work addressing variety of time-series classification techniques as applied on multiple domains is described as follows. Lines et al., 2012, propose a shapelet transform for time-series classification. Their implementation includes the development of a caching algorithm to store shapelets, and to apply a parameter-free cross-validation approach for extracting the most significant shapelets. Experiments included the transformation of 26 data-sets to demonstrate that a *C4.5* decision tree classifier trained with transformed data is competitive with an implementation of the original shapelet decision tree of Ye and Keogh, 2009. Lines et al., 2012, demonstrate that the filtered data can be applied also to non-tree based classifiers to achieve improved classification performance, while maintaining the interpretability of shapelets. Another signal classification approach is presented in Povinelli et al., 2004—the approach is based upon modeling the dynamics of a system as they are captured in a reconstructed phase space. The modeling is based on full covariance Gaussian Mixture Models of time domain signatures. Three data-sets were used for validation, including motor current simulations, electrocardiogram recordings, and speech waveforms. The approach is different than other signal classification approaches (such as linear systems analysis using frequency content and simple non-linear machine learning models such as artificial neural networks). The results demonstrate that the proposed method is robust across these diverse domains, outperforming the time delay neural network used as a baseline. Using artificial neural networks for classifying time-series, however, as described in (Haselsteiner and Pfurtscheller, 2000) proved to be robust—the authors address classification of electroencephalograph (EEG) signals using neural networks. The paper compares two topologies of neural networks. Standard multi-layer perceptrons (MLPs) are used as a method for classification, and are compared to finite impulse response (FIR) MLPs, which use FIR filters instead of static weights to allow temporal processing inside the classifier. Experiments with three different subjects demonstrate the higher performance of the FIR MLP compared with the standard MLP. Another example for using supervised learning (recurrent neural networks) for classifying time-series is provided as in (Hüsken and Stagge, 2003).

Perng et al., 2000, propose the *Landmark Model* for similarity-based pattern querying in time-series databases - a model of similarity that is consistent with human intuition and episodic memory. *Landmark Similarity* measures are computed by tracking different specific subsets of features of landmarks. The authors report on experiments using 10-year closing prices of stocks in the *Standard & Poor 500* index. Nguyen et al., 2011, propose an algorithm called *LCLC (Learning from Common Local Clusters)* to create a classifier for time-series using limited labeled positive data and a cluster chaining approach to improve accuracy. The authors compare the *LCLC* algorithm with two existing semi-supervised methods for time-series classification: Wei's method (Wei and Keogh, 2006), and Ratana's method (Ratanamahatana and Wanichsan, 2008). To demonstrate the superiority of *LCLC*, the authors used five data-sets of time-series (Wei, 2007; Keogh, 2008).

Jović et al., 2012, examined the use of decision tree ensembles in biomedical time-series classification. Experiments performed focused on biomedical time-series data-sets related to cardiac disorders, demonstrated that the use of decision tree ensembles provide superior results in comparison with support vector machines (SVMs). In particular, *AdaBoost.M1* and *MultiBoost* algorithms applied to *C4.5* decision tree found as the most accurate. Esmael et el., 2012, propose a feature-based classification approach to classify time-series generated by drilling rig sensors. The approach is based on two phases: representation and classification. The authors concluded that memory-based classifiers improve the classification accuracy of time-series significantly. Kampouraki et al., 2009, describe an implementation of SVMs to classify heartbeat time-series. The authors report on the superiority of using SVMs in comparison with neural network-based classification approaches. Experimental data included two data-sets of ECG (Electrocardiography) recordings measured during several-hour time periods each: the first dataset consists of long-term ECG recordings of young and elderly healthy subjects. The second dataset consists of long-term ECG recordings of normal subjects and subjects suffering from coronary artery disease. Examples for additional significant work in the domain of time-series classification is presented in Zhang et al., 2008; Luca and Zuccolotto, 2011; Sugimura and Matsumoto, 2011, Xi et al., 2006; Nguyen et al., 2012; and Wiens et al., 2012.

The paper describes a search engine system that is based on time-series classification to identify similarities among time-series representing monthly returns of hedge funds. The search approach allows an investor to instantaneously identify candidate hedge funds to diversify his or her portfolio eliminating the complexity of the user experience involved in existing hedge funds' selection database systems. The approach is based on applying new machine learning methods to classify long time-series of monthly returns representing hedge funds. The *self-labeling* and the *signal composition* methods allow evaluating the level of similarity between the behaviors of time-series for extended periods of time.

The structure of the paper is as follows: after introducing the financial search system in Section 2, the time-series classification methods are detailed in Section 3. Section 4 presents experiments to validate the classification accuracy. Discussion and conclusions are provided in Sections 5 and 6, respectively.

## 2. A Search Engine System

### 2.1 System Architecture

As shown in figure 1, the system architecture employs a client computer and a classifying server. The client computer facilitates a user to specify a hedge fund via a user interface presented on a display. The hedge fund specified by the user is sent to a classifying database, operatively coupled with classifying server, in a textual format. The classifying database contains several tables including data structures such as a table of classification results, and a table that contains information about a large collection of hedge funds, e.g., monthly returns. Once a user-request is received by the classifying server, the classifying server processes the user-request and sends processed classification results back to the client computer.

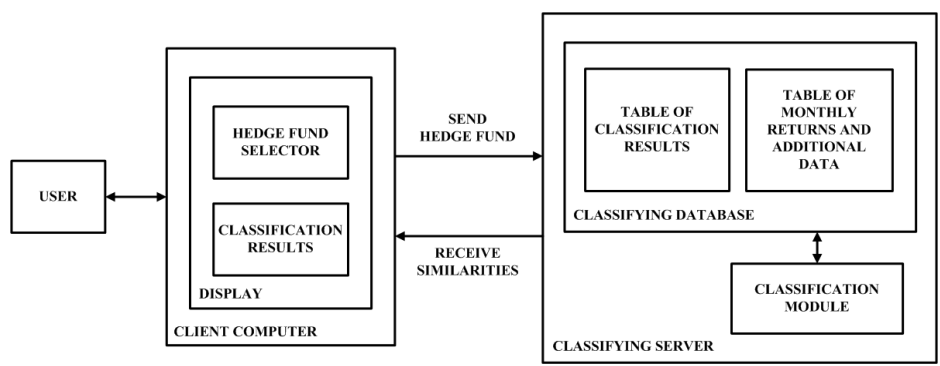

*Figure 1.* System architecture.

In response to receiving the processed classification results, i.e., a list of hedge funds, the client computer provides on the display results that include the representation of the hedge funds, wherein the hedge funds behave most similarly to the specified hedge fund during a pre-defined time range (e.g., a decade). In addition, the client computer receives additional details associated with the specified hedge fund and the similarly behaving hedge funds. Such additional details include: *Category* (e.g., Emerging Market Equity, Fund of Funds, Multi-strategy), *Contact Information*, *Net Assets*, *Minimum Initial*,

*Management Fee*, *Performance Fee*, *Deferred Load*, *Redemption Fee*, *Base Currency*, *Total Returns* (e.g., 1 Month, 3 Months, 6 Months, 3 Years, 5 Years), and *Domicile* (e.g., United States, Switzerland, Cayman Islands).

The table of classification results (figure 1) is formed by applying classification procedures. A classification module includes the classification procedures and is a component of the classifying server. The classification module generates the content of the table of classification results based on monthly returns of the hedge funds. The classification module and the classification procedures will be described in more detail further in the text referring to figure 4 and expressions 3.1 - 3.4.

## 2.2 Human-Computer Interaction

Figures 2 & 3 provide the flow diagram and the user interface for employing the financial search engine, respectively. As shown, the user specifies a hedge fund which is sent to the classifying database, coupled to operate with classifying server. A list of hedge funds and additional details associated with the hedge funds are received from the classifying database. The list contains hedge funds that found to have similar behavioral patterns to the hedge fund specified. The list is sorted according to a level of similarity criterion and presented at the user interface. Additional financial details associated with the hedge funds are presented.

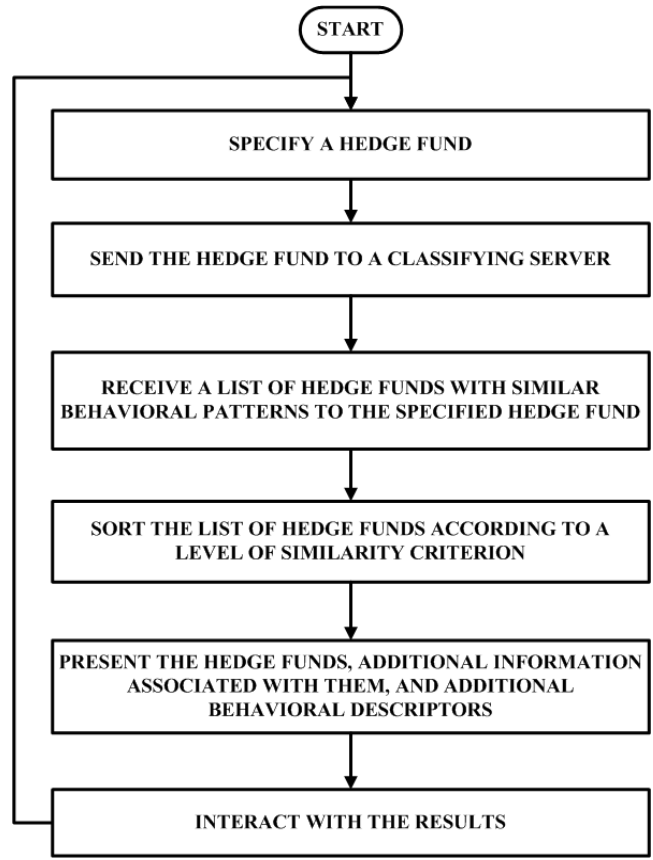

*Figure 2.* Flow diagram for employing the financial search engine.

(a) A user specifies a hedge fund and receives similarly behaving hedge funds.

Kartoun U., White Paper: A Method for Comparing Hedge Funds, 2013.

(b) An expanded view at one of the similarly behaving hedge funds identified.

*Figure 3.* A search engine system for hedge funds - user interface.

As an example, a user may specify *Ardsley Partners Renewable Energy LP* (Category: Global Equity, Domicile: United States). Immediately acquired from the database a list of hedge funds with similar behavior to the specified hedge fund during the pre-defined time period, January 2000 - August 2010. The most similar hedge funds found are shown in Table 1 sorted in a descending order according to a similarity criterion. As seen from Table 1, two hedge funds that were identified as behaving similarly to *Ardsley Partners Renewable Energy LP* demonstrate the ability of the classification methods and system to classify hedge funds from different categories. Further, *FPP Emerging Markets Limited* is from a different country than *Ardsley Partners Renewable Energy LP*.

*Table 1.* An example for two hedge funds acquired for *Ardsley Partners Renewable Energy LP* (January 2000 - August 2010)

| Hedge Fund Name | Category | Domicile | Benefits in comparison with Ardsley Partners Renewable Energy LP |
|---|---|---|---|
| Arrow Partners LP | Corporate Actions | United States | Higher 1-Month Return<br>Higher 3-Month Return |
| FPP Emerging Markets Limited | Emerging Market Equity | Cayman Islands | Lower Performance Fee<br>Higher 1-Month Return<br>Higher 3-Month Return<br>Higher 6-Month Return<br>Higher 1-Year Return<br>Higher Sharpe Return |

Next to each similarly behaving hedge fund presented also one or more indicators specifying the hedge fund's superiority in comparison with *Ardsley Partners Renewable Energy LP*. An indicator is an expression that represents a benefit between the specified hedge fund and each of the hedge funds found. Examples for such indicators include:

- An expression that represents the difference in fees between the specified hedge fund and the acquired similarly behaving hedge fund, e.g., *Lower Performance Fee*,

- An expression that represents the difference in return between the specified hedge fund and the acquired similarly behaving hedge fund, e.g., *Higher 6-Month Return*, and,

- An expression that represents the difference in risk between the specified hedge fund and the acquired similarly behaving hedge fund, e.g., *Higher Sharpe Ratio*.

Additional characteristics and the characteristics' corresponding values associated with the hedge funds found, as mentioned in Section

2.1 are also presented next to the specified hedge fund and next to each result.

## 2.3 Classification Module

To classify hedge funds, the classification module is used as shown in figure 4. The classification module is facilitated to perform a method that generates classification results stored in the classifying database (figure 1). The classification method is applied on all of the patterns of the hedge funds considered. The available patterns are of monthly returns of hedge funds operated during approximately one decade (January 2000 - August 2010). The monthly returns are served along with the outcome of a self-labeling method (expressions 3.1 - 3.4), as input for a decision tree learning algorithm as described in more detail in Section 3.1.

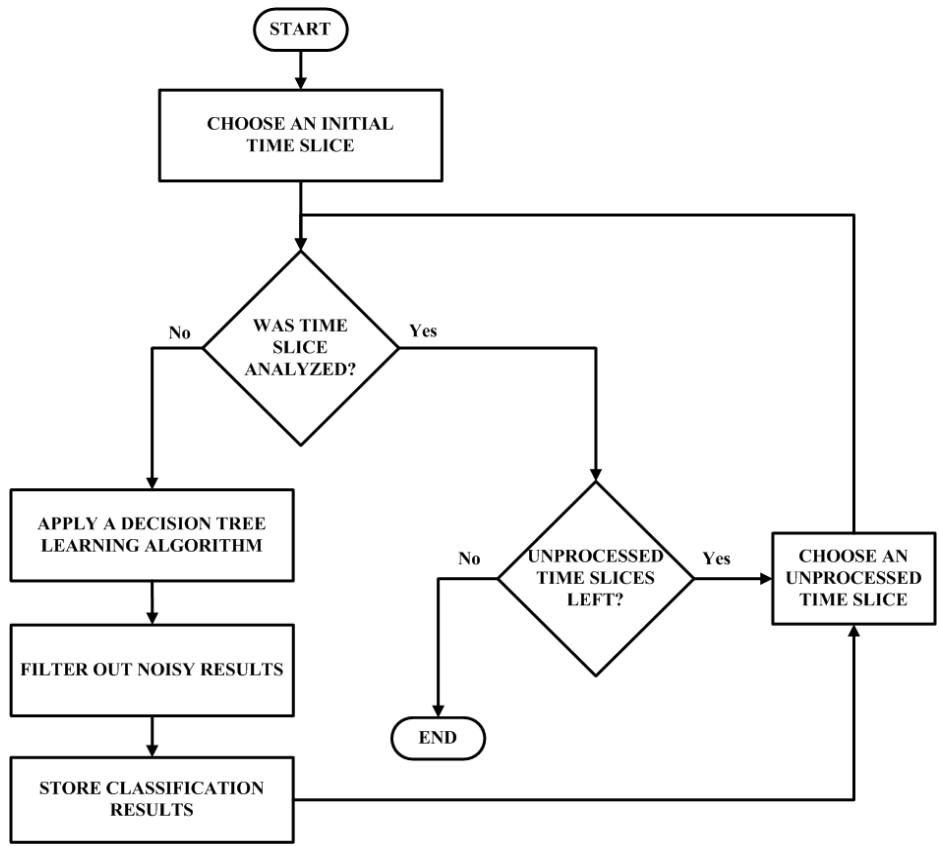

*Figure 4.* Classification method.

## 3. Time-Series Classification

### 3.1 Self-Labeling

Assume $H_1, H_2, H_i, ... H_m$ are $m$ hedge funds considered for classification during a time range that includes $n$ time-steps (*e.g.*, a one time-step equals to one month). Each hedge fund $H_i$ is associated with a vector of returns in which vector of returns is denoted as $H[R]$. For a hedge fund $H_i$ the vector of returns is presented as follows:

$$H_1[R_1, R_2, R_j, ... R_n]$$
$$H_2[R_1, R_2, R_j, ... R_n]$$
$$...$$
$$H_i[R_1, R_2, R_j, ... R_n]$$
$$...$$
$$H_m[R_1, R_2, R_j, ... R_n]$$

3.1

Expressions 3.2 represent slicing the vectors of returns (expressions 3.1) to collections of six-month each ($h = 6$):

$$H_1[R_1,R_2,R_3,R_4,R_5,R_6]_1, H_1[R_7,R_8,R_9,R_{10},R_{11},R_{12}]_2, ... H_1[R_{n-5},R_{n-4},R_{n-3},R_{n-2},R_{n-1},R_n]_k$$
$$H_2[R_1,R_2,R_3,R_4,R_5,R_6]_1, H_2[R_7,R_8,R_9,R_{10},R_{11},R_{12}]_2, ... H_2[R_{n-5},R_{n-4},R_{n-3},R_{n-2},R_{n-1},R_n]_k$$
$$...$$
$$H_i[R_1,R_2,R_3,R_4,R_5,R_6]_1, H_i[R_7,R_8,R_9,R_{10},R_{11},R_{12}]_2, ... H_i[R_{n-5},R_{n-4},R_{n-3},R_{n-2},R_{n-1},R_n]_k$$
$$...$$
$$H_m[R_1,R_2,R_3,R_4,R_5,R_6]_1, H_m[R_7,R_8,R_9,R_{10},R_{11},R_{12}]_2, ... H_m[R_{n-5},R_{n-4},R_{n-3},R_{n-2},R_{n-1},R_n]_k$$

3.2

where the size of the total time range of $n$ time-steps, also equals to $k$ time slices each of length of six monthly returns ($h = 6$). The following representation, for example, is considered for the first time slice ($k = 1$):

$$H_1[R_1, R_2, R_3, R_4, R_5, R_6]_1$$
$$H_2[R_1, R_2, R_3, R_4, R_5, R_6]_1$$
$$...$$
$$H_i[R_1, R_2, R_3, R_4, R_5, R_6]_1 \quad\quad 3.3$$
$$...$$
$$H_m[R_1, R_2, R_3, R_4, R_5, R_6]_1$$

In the classification problem considered here no labels are available for the time-series and there is no information on how to refer to a set of values associated with a certain time-series. As such, a numerical value representing each time-series is generated and assigned as the label of the time-series. The numerical value label denoted as $LH_i$ is calculated for each time-series in a time slice:

$$LH_1 = \sum_{l=1}^{h} H_1(R_l)$$
$$LH_2 = \sum_{l=1}^{h} H_2(R_l)$$
$$...$$
$$LH_i = \sum_{l=1}^{h} H_i(R_l) \quad\quad 3.4$$
$$...$$
$$LH_m = \sum_{l=1}^{h} H_m(R_l)$$

The representation of self-labeling as shown in 3.4 expressions facilitates the application of supervised learning methods on unlabeled data sets. This is achieved by providing a supervised learning classification algorithm with pairs of adjusted representations of original time-series (as shown as an example for $k=1$ in 3.3 expressions) and the adjusted representations' corresponding self-generated label (3.4 expressions).

## 3.2 Decision Tree Learning

The procedure described through expressions 3.1 - 3.4 is applied by acting several tables stored in a classifying database (figure 1). Values according to expressions 3.3 and their corresponding labels as in expressions 3.4 are served as an input for a standard supervised learning algorithm. The supervised learning algorithm used in this paper is a decision tree algorithm. For each time slice, a decision tree is generated. An example for a partial representation of a decision tree is shown in figure 5. A decision tree is a data structure that consists of branches and leaves. Leaves (also denoted as "nodes") represent classifications, and branches represent conjunctions of features that lead to those classifications. Each node has a unique title to distinguish the node from other nodes that the tree composed of. A node contains two or more records. Each record represents a hedge fund, its feature values (3.3 expressions) and its predictor value (3.4 expressions). The fewer hedge fund records in a node (the minimum is two), the less the node varies, i.e., a node with fewer records is more likely to represent a better classification between the hedge funds that the node contains.

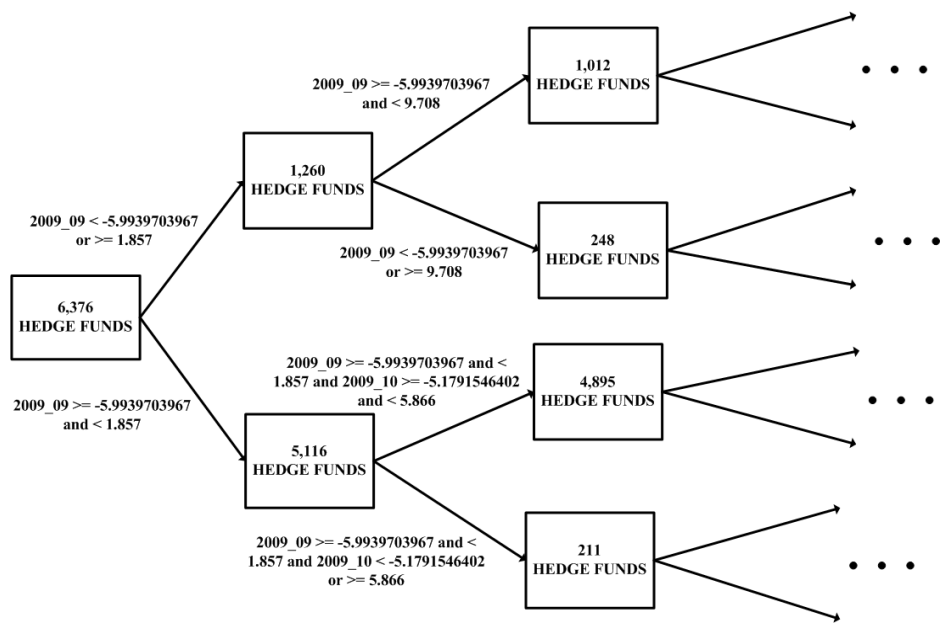

*Figure 5.* A partial representation of a decision tree for July 2009 - December 2009 (1722 nodes).

   The number of nodes in a generated tree depends on the duration of the time slice and the number of hedge funds considered. The classification accuracy of the algorithm depends on its input parameters. Parameters for a decision tree algorithm include complexity penalty, to control the growth of the decision tree, and minimum support, to determine the minimal number of leaf cases required to generate a split. Setting the desired values for the decision tree algorithm parameters depends on the tradeoff between classification accuracy and computational speed. Classifying with perfect or close to perfect accuracy thousands of hedge funds may require extended periods of time to apply a decision tree algorithm. To reduce the calculation time, the growth of the decision tree is controlled by increasing the complexity penalty level (this decreases the number of splits) and by increasing the level of minimum support. On one hand, controlling the growth of the tree improves computation performance. On the other hand, controlling the growth of the tree may affect classification accuracy. As shown in figure 4, a filtering procedure is applied to each decision tree generated to partially overcome this and to avoid identifying groups of hedge funds that behave differently from each other but are still classified as similar. For each tree, the predictor value of each hedge fund in a node is compared with the other predictors of the hedge funds present in the node. If the variability of predictors found in a node is above a pre-defined threshold, then the node is considered a noisy/inaccurate classification, i.e., the node is pruned.

11,312 hedge funds are considered for classification. The total time range for classification is 128 months (January 2000 - August 2010). For most of the hedge funds considered monthly return information was available for the entire time range, however, for certain hedge funds data was available only partially (e.g., returns for *3A Asia Fund CHF A* are only available from July 2007). For each 6-month collection of returns from 2000 to 2009 and for January 2010 - August 2010 (8 months), a decision tree based classification was performed - each such classification results a decision tree data structure (total of 21 decision trees). For the amount of data considered here, a typical size for one decision tree is in the range of approximately 200 to 2,300 nodes. The decision tree includes a main node that contains all considered hedge funds[2]. The decision tree algorithm generates rules. The rules are based on the monthly returns of the hedge funds. Some nodes in the tree split to two sub-nodes, i.e., children, and other nodes do not. A split, if occurs, is based on the generated rules and separates a group of hedge funds to two smaller groups. For example, for the main node that consists of 6,376 hedge funds, two rules were generated:

- 2009_09 < -5.9939703967 or >= 1.857

- 2009_09 >= -5.9939703967 and < 1.857

while "2009_09" stands for September 2009. Similarly, other generated rules split nodes across the tree.

The decision tree classification results for a time slice considered, excluding noisy data, are stored in table of classification results of classifying database of the classifying server (figure 1). Table 2 is an exemplary partial representation of the table of classification results for a period of six months (one time slice represented by one decision tree). For the amount of data considered here, the number of records representing the nodes of one decision tree classification results is in the range of approximately 800 to 8,000 records.

---

[2] The figure presents a decision tree that includes 6,376 of the entire set of 11,312 time-series considered. The reason for that—time-series that do not contain all monthly returns for the period considered (typically six values) are omitted from the analysis for that time range.

The procedure repeats itself with the next time slice until all of the 21 time slices of January 2000 - August 2010 are processed and decision trees are created for them and added in a tabular format to the table of classification results. For the amount of data considered here, the number of records in the table is approximately 75,000. The classification method (figure 4) is performed only once. When the classification method is completed and the table of classification results is created, users may query the table using the client computer as previously described within the context of figures 1 through 3.

*Table 2.* An example for a tabular representation of one decision tree classification.

| Node Name | Hedge Fund Name |
|---|---|
| A | Skandia Global Hedge Fund |
| A | Andorfons Alternative Premium Acc |
| A | Akros Absolute Ret Acc |
| B | GAM Multi-North America Inc. USD Special |
| B | ADI GLOBAL REGA A |
| B | Man Gbl Str Div Ser 2 Ltd-CHF Cap |
| C | ABN AMRO Alt Inv ARAF V450 I |
| C | SC Trend EUR |
| … | … |

### 3.3 Signal Composition

To receive similarities for a hedge fund, a signal composition method also denoted as time-series composition method is applied (figure 6). Consider a hedge fund and a time range specified by the user[3]—the hedge fund is denoted as $H$ and the time range is represented by a set of $t$ decision trees each representing one time slice classification.

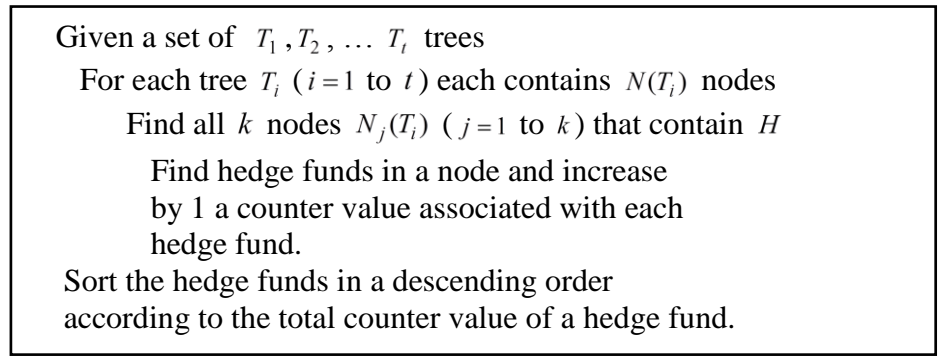

Given a set of $T_1, T_2, \ldots T_t$ trees
  For each tree $T_i$ ($i = 1$ to $t$) each contains $N(T_i)$ nodes
    Find all $k$ nodes $N_j(T_i)$ ($j = 1$ to $k$) that contain $H$
      Find hedge funds in a node and increase
      by 1 a counter value associated with each
      hedge fund.
  Sort the hedge funds in a descending order
  according to the total counter value of a hedge fund.

*Figure 6.* Signal composition method.

To rank the level of similarity between a specified hedge fund to the hedge funds found according to the signal composition (figure 6), the counter value is used—the higher the counter value for a hedge fund,

---

[3] The length of a time range could also be determined in advance without letting the user specify it (e.g., one decade).

the more similarly behaving the hedge fund is to the specified hedge fund.

## 4. Experiments

To evaluate the accuracy of the search engine, *Ardsley Partners Renewable Energy LP*, mentioned in Section 2.2 is discussed first. Two of the most similarly behaving hedge funds identified by using the methods discussed in Section 3; *self-labeling*, *decision tree learning*, and *signal composition*, are presented in Table 3. The numerical values presented in the table are *Pearson Product-Moment Correlation Coefficient* values between the hedge funds ($r$). High values of $r$ indicate a high correlation between the time-series representing the hedge funds. As seen from Table 3, strong correlations, i.e., high $r$ values, of 0.77 and 0.84 were calculated for *Arrow Partners LP*, and *FPP Emerging Markets Limited*, respectively (see figure 7).

*Table 3.* Correlations between Ardsley Partners Renewable Energy LP and hedge funds identified as similar (January 2008 - August 2010).

|  | **Ardsley Partners Renewable Energy LP** | **Arrow Partners LP** | **FPP Emerging Markets Limited** |
|---|---|---|---|
| **Ardsley Partners Renewable Energy LP** | 1.00 | 0.77 | 0.84 |
| **Arrow Partners LP** | 0.77 | 1.00 | 0.79 |
| **FPP Emerging Markets Limited** | 0.84 | 0.79 | 1.00 |

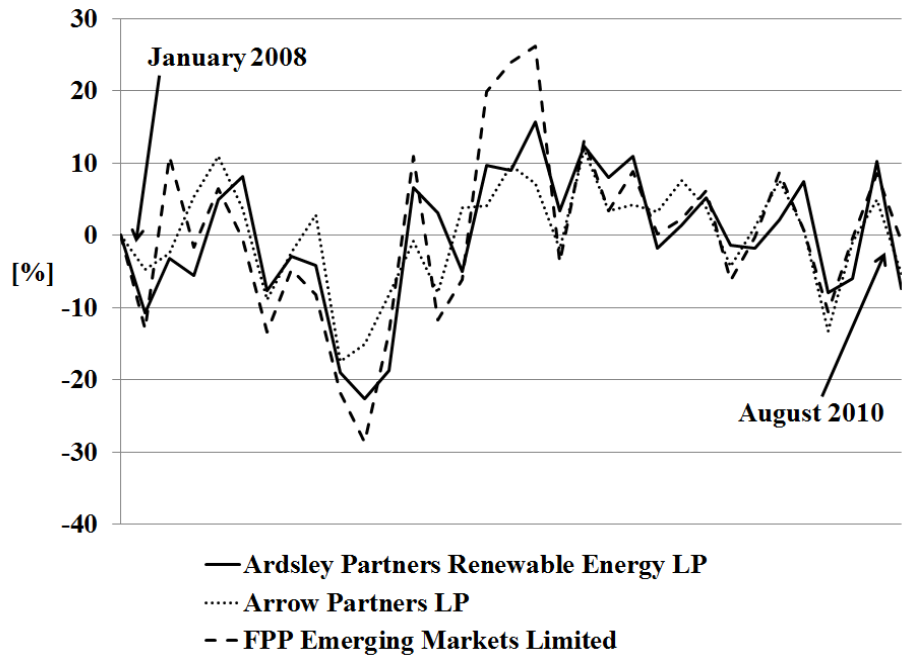

*Figure 7.* Similarly behaving hedge funds to Ardsley Partners Renewable Energy LP.

In a second experiment, an arbitrary hedge fund was picked: *Bridgewater Pure Alpha Trad Ltd*. Two of the most similarly behaving hedge funds identified are presented in Table 4. The high values of *r* presented in the table indicate a high correlation between the time-series representing the hedge funds. As seen from Table 4, positive correlations, i.e., high *r* values, of 0.99 and 0.35 were calculated for *Global Pure Alpha Fund Class B*, and *Zephyr Commodity Fund CHF*, respectively (see figure 8).

*Table 4.* Correlations between Bridgewater Pure Alpha Trad Ltd and hedge funds identified as similar (October 2006 - February 2010).

|  | **Bridgewater Pure Alpha Trad Ltd** | **Global Pure Alpha Fund Class B** | **Zephyr Commodity Fund CHF** |
| --- | --- | --- | --- |
| **Bridgewater Pure Alpha Trad Ltd** | 1.00 | 0.99 | 0.35 |
| **Global Pure Alpha Fund Class B** | 0.99 | 1.00 | 0.35 |
| **Zephyr Commodity Fund CHF** | 0.35 | 0.35 | 1.00 |

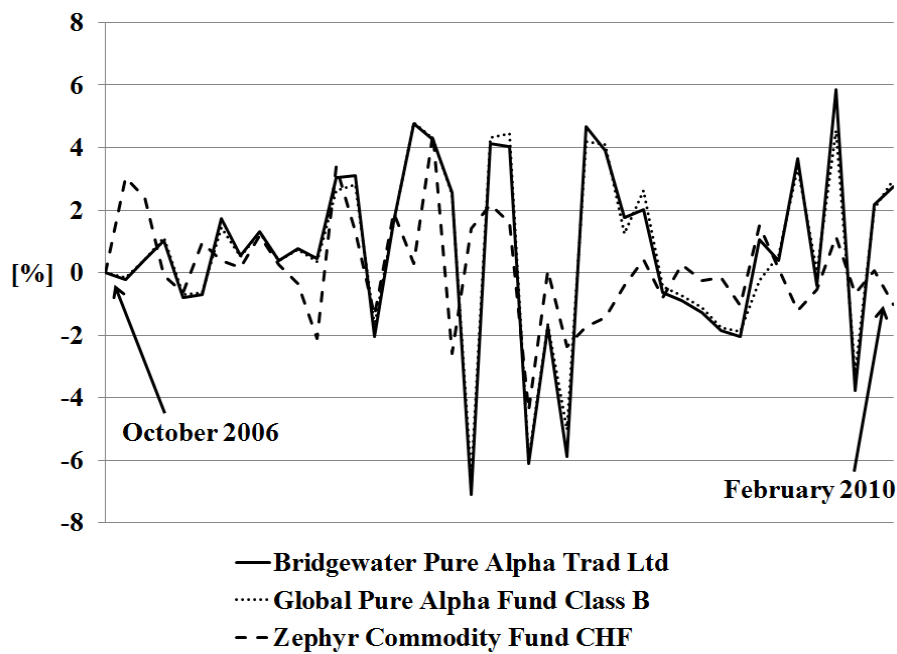

*Figure 8.* Similarly behaving hedge funds to Bridgewater Pure Alpha Trad Ltd.

Kartoun U., White Paper: A Method for Comparing Hedge Funds, 2013.

In a third experiment, an arbitrary hedge fund was picked: *HSBC Distressed Oppor Fund USD*. Two of the most similarly behaving hedge funds identified are presented in Table 5. The high values of $r$ presented in the table indicate a high correlation between the time-series of the hedge funds. As seen from Table 5, positive correlations, i.e., high $r$ values, of 0.61 and 0.56 were calculated for *Advent Convertible Arbitrage Fund*, and *Crosslink Emerging Growth Fund, L.P.*, respectively (see figure 9).

*Table 5.* Correlations between HSBC Distressed Oppor Fund USD and hedge funds identified as similar (March 2006 - August 2010).

|  | HSBC Distressed Oppor Fund USD | Advent Convertible Arbitrage Fund | Crosslink Emerging Growth Fund, L.P. |
|---|---|---|---|
| **HSBC Distressed Oppor Fund USD** | 1.00 | 0.61 | 0.56 |
| **Advent Convertible Arbitrage Fund** | 0.61 | 1.00 | 0.53 |
| **Crosslink Emerging Growth Fund, L.P.** | 0.56 | 0.53 | 1.00 |

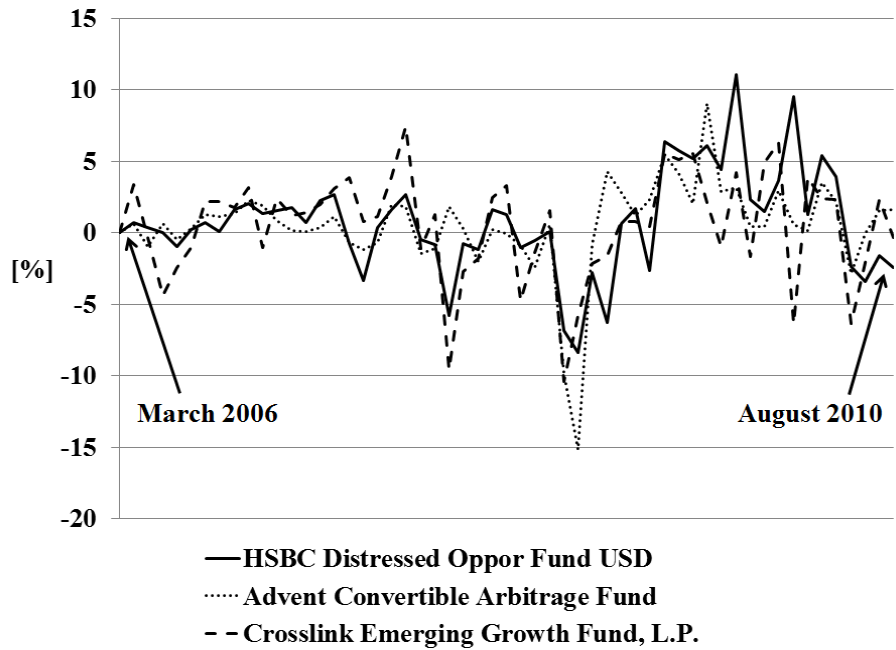

*Figure 9.* Similarly behaving hedge funds to HSBC Distressed Oppor Fund USD.

The results shown in tables 3 through 5 and in figures 7 through 9 reflect the ability of the presented classification methods and the financial search engine to identify similarities among hedge funds, i.e., pairs of hedge funds with positive high correlation, regardless of category and location.

## 5. Discussion

Providing classifications for signals or time-series is well discussed in the literature; however, what is significant in the paper is the search approach that allows an investor to instantaneously identify candidate hedge funds to diversify his or her portfolio. The approach is based on applying new machine learning methods to classify long time-series of representing monthly returns. The *self-labeling* and the *signal composition* methods as described through 3.1 - 3.4 expressions allow evaluating the level of similarity between the behaviors of time-series for extended periods of time. The main objective to use time slices is to reduce the computation complexity—in practice, using too large number of input features in a classification algorithm may result unfeasible processing times. Splitting a collection of long time-series into short time slices, performing classification for the shorter time slices separately and then applying the proposed signal composition method, provides feasible processing times. Another reason to use time slices is the improved classification accuracy for certain problems.

## 6. Conclusion

The paper presents new machine learning methods: *signal composition*, which classifies time-series regardless of length, type, and quantity; and *self-labeling*, which is a supervised-learning enhancement. The methods were implemented on a financial use case as a search engine. The methods and the system were used to classify time-series of 11,312 hedge funds operated for approximately one decade (January 2000 - August 2010). The search engine allows a user to specify a particular hedge fund and a time range to query a database to receive in real-time a list of hedge funds that behave similarly to the particular hedge fund and the time range provided. The presented search approach of a cross-category and cross-location classification assists the user to make diversification decisions in his or her portfolio. The human-computer interaction financial search approach is unique: specifying a hedge fund and a time range, receiving similarities, and presenting benefits next to each result in comparison with what specified. The search approach and methods could be used to develop stand-alone financial decision support systems. Alternatively, the methods could be embedded in existing portfolio management systems and financial screeners using cloud-computing technology. The methods could also be integrated with portfolio evaluation and risk management systems such as described in Liu et al., 2006; Yang, 2010; Konno and Yamazaki, 1991; and Lukyanitsa et al., 2009.

Although the paper describes an implementation that relates to classification of hedge funds, the methods described, i.e., *self-labeling* and *signal composition*, could be implemented on other use cases. For example, the proposed methods and system could be applied to mutual funds, exchange-traded-finds and stocks, or to series of non-financial behavioral patterns such as seismic or bio-medical patterns.